\documentclass[aps,prl,showpacs,twocolumn]{revtex4}

\usepackage{amssymb}
\usepackage{graphicx}
\usepackage{amsmath}
\usepackage{amsbsy}
\usepackage{amsthm}
\usepackage{bbm}
\usepackage{bm}
\usepackage{epsfig}
\newcommand{\id}{\mathbbm{1}}

\newcommand{\ra}{\rangle}
\newcommand{\la}{\langle}
\newcommand{\be}{\begin{equation}}
\newcommand{\ee}{\end{equation}}
\newcommand{\beq}{\begin{eqnarray}}
\newcommand{\eeq}{\end{eqnarray}}
\newcommand{\bra}[1]{\ensuremath{\langle #1 |}}
\newcommand{\ket}[1]{\ensuremath{| #1 \rangle}}

\begin{document}

\title{Detection of high-dimensional genuine multi-partite entanglement of mixed states}

\author{Marcus Huber, Florian Mintert}
\affiliation{Faculty of Physics, University of Vienna, Boltzmanngasse 5, 1090 Vienna, Austria}
\affiliation{Institute of Physics, Albert-Ludwigs University of Freiburg, Hermann-Herder-Str. 3, 79104 Freiburg Germany}

\author{Andreas Gabriel, Beatrix C. Hiesmayr}
\affiliation{Faculty of Physics, University of Vienna,
Boltzmanngasse 5, 1090 Vienna, Austria}

\begin{abstract}
We derive a general framework to identify genuinely multipartite entangled mixed quantum states in arbitrary-dimensional systems and show in exemplary cases that the constructed criteria are stronger than those previously known. Our criteria are simple functions of the given quantum state and detect genuine multipartite entanglement that had not been identified so far. They are experimentally accessible without quantum state tomography and are easily computable as no optimization or eigenvalue evaluation is needed.
\end{abstract}


\maketitle

Many-particle entanglement is a striking feature of quantum many-body systems.
Entanglement was first recognized as a curiosity of quantum mechanics
because it gives rise to seemingly nonlocal correlations of measurement results of distant observers.
Whereas the central role of many-body entanglement for various applications of quantum information processing (e.g. \cite{qc})
is undoubted,
its role in {\it e.g.} quantum phase transitions (e.g. \cite{phase}) or ionization processes is still debated (e.g. \cite{helium}),
and questions concerning {\it e.g.} its potential assistance to the astonishing transport efficiency of biological compounds (e.g. \cite{bio}) are still essentially open.

To answer such questions we need reliable techniques to characterize entanglement properties of general quantum states.
However, even the conceptually rather simple question `Is a given quantum state entangled or not{?}' is in general unanswered so far.
It is usually addressed by means of separability criteria, which work very well in many
cases, but are far from perfect \cite{horodeckiqe}.
Even more challenging is the detection of genuine multipartite entanglement, which has already been intensely
studied (see for example \cite{horodeckicrit, wocjancrit, yucrit,hassancrit}),
but still has not yielded satisfying results.
Vast areas of the considered state-spaces are still widely unexplored due
to the lack of suitable tools for detecting and characterizing entanglement.

The central difficulty arises from the complicated structure of multipartite entangled states:
even states that do not  separate into blocks of subsystems that are {\em not} entangled with each other
are not necessarily genuinely $n$-body entangled.
Recently, inequalities to identify genuinely $n$-body entangled states have been proposed based on nonlinear functions of matrix-elements \cite{seevinckcrit}.
Although these new criteria are promising in the sense that they allow us to characterize states as entangled that can not be detected with the standard criteria,
it is also evident that the characterization of entangled states will not be facilitated by a huge set of separability criteria unless we have a systematic way
to construct and understand these criteria.
Here, we present a very general, systematic approach to construct such criteria,
and show that our newly constructed criteria are stronger than all formerly known ones.
In particular, all these criteria apply to systems of arbitrarily many subsystems of arbitrary finite dimensions.

In more detail, we derive\\
- an $m$-linear inequality ~\eqref{eq:bipartite} and its bilinear version ineq.\eqref{eq:bilin} to detect bipartite entanglement.
Based on this, we derive\\
- a general framework to obtain bilinear inequalities ~\eqref{eq:genmulti} which characterize genuine multipartite entanglement and\\
- construct a particularly strong criterion, i.e. ineq.~\eqref{eq:genmulti2}, for which the efficiency is demonstrated in the consecutive examples.

A pure $n$-partite state $\ket{\Psi}$ is called $k$--\emph{separable} if it can be written as a product
\cite{horodeckiqe}
\begin{eqnarray}
|\Psi\rangle=|\phi_1\rangle\otimes|\phi_2\rangle\otimes\cdots\otimes|\phi_k\rangle\ ,
\label{eq:ksep}
\end{eqnarray}
of $k$ states $\ket{\phi_i}$ each of which corresponds to a single subsystem or a group of subsystems.
If there is no such form with at least two factors, then $\ket{\Psi}$ is considered genuinely $n$-partite entangled.
On the level of pure states the question of $k$-separability can be answered in a straight forward fashion by means of separability criteria for bipartite systems,
simply by considering all segmentations of the $k$-partite system into two parts.
However, the same question becomes significantly more difficult to answer for mixed states $\varrho$:
here, a state is considered genuinely $k$-partite entangled
if {\em any} decomposition into pure states
\be
\varrho=\sum_ip_i\ket{\psi_i}\bra{\psi_i} \ ,
\ee
with probabilities $p_i>0$ contains at least one genuinely $k$-partite entangled component.
Therefore, a mixed state can still be partially separable, even if the $k$ subsystems can not be split into two groups that are not entangled with each other.
Consider for instance the tri-partite state
\be \rho_{bisep}=\sum_j p_j
\rho_{AB}^j\otimes\rho_C^j+\sum_j q_j
\rho_{AC}^j\otimes\rho_B^j+\sum_j r_j
\rho_{BC}^j\otimes\rho_A^j\;,\nonumber \ee 
Here the two-body states $\rho^j_{AB}$, $\rho^j_{BC}$ and $\rho^j_{AC}$ describe entangled states. 
Even though there is no bipartite splitting with respect to which the state $\rho$ is separable,
it is considered biseparable since it can be prepared through a statistical mixture of bipartite entangled states.

To be certain that some state is really genuinely $n$-body entangled, we thus have to make sure that there is no pure state decomposition with only at least partially entangled components.
Since this reduces to the problem of deciding whether each of such pure state components is at least biseparable, let us first introduce a suitable criterion for biseparability,
which then will turn out to be the central building block for the subsequent generalization to genuine many-body entanglement.
What we employ here, are $m$-linear functions of a quantum state $\varrho$ on $\mathcal{H}_A\otimes\mathcal{H}_B$
that can be expressed in terms of the $m$-fold tensor product $\varrho^{\otimes m}$ of the density matrix $\varrho$
acting on the $m$-fold tensor product space $(\mathcal{H}_A\otimes\mathcal{H}_B)^{\otimes m}$.
As it is shown at the end of our letter, any separable state $\varrho_s$ satisfies
\beq
\sqrt{|\Re e(\bra{\Phi}(\id\otimes\Pi_B)^\dagger\varrho_s^{\otimes m}(\Pi_A\otimes\id)\ket{\Phi})|}\le\sqrt{\bra{\Phi}\varrho_s^{\otimes m}\ket{\Phi}}\ ,
\label{eq:bipartite}
\eeq
for any positive integer $m$,
where $\ket{\Phi}$ is any fully separable state of the $m$-tupled system, {\it i.e.} $\ket{\Phi}$ factorizes into $2m$ single-body states. 
$\Pi_{A}$ is the cyclic permutation operator acting on $\mathcal{H}_{A}^{\otimes m}$, {\it i.e.}
\be
\Pi_A\ket{\varphi_1}\otimes\ket{\varphi_2}\otimes\hdots\otimes\ket{\varphi_m}=\ket{\varphi_2}\otimes\ket{\varphi_3}\otimes\hdots\otimes\ket{\varphi_m}\otimes\ket{\varphi_1} \ ,
\ee
and $\Pi_B$ is defined analogously for subsystem $B$.
In our following extension to multipartite systems, we will content ourselves with the bilinear case
$m=2$, as it is already very powerful in detecting entanglement and ineq.~\eqref{eq:bipartite} takes the rather simple form
\be
|\langle il|\rho|kj\rangle|-\sqrt{\langle ij|\rho|ij\rangle\langle kl|\rho|kl\rangle}\leq 0\label{eq:bilin}\ ,\tag{I}
\ee
which corresponds to the choice $\ket{\Phi}=\ket{ijkl}$.\\
For our following generalization of ineq.~\eqref{eq:bipartite} to the multipartite case
we will consider all $(2^{n-1}-1)$ different partitions of an $n$-partite systems into two subsystems,
because a mixed state is biseparable exactly if there is a decomposition into
pure states each of which is separable with respect to some partition.
The fictitious subsystems will be labeled $A_i$ and $B_i$ ($i=1,\hdots,2^{n-1}-1$) in the following.
Introducing the {\em global} permutation operator ${\bf \Pi}$ which performs simultaneous permutations on all subsystems,
we can formulate now the generalization of ineq.~\eqref{eq:bipartite} to multipartite systems:
\be
\tag{II}
\sqrt{\langle\Phi|\rho^{\otimes 2}{\bf \Pi}|\Phi\rangle}-\sum_i\sqrt{\langle\Phi|{\cal P}_i^\dagger\rho^{\otimes 2}{\cal P}_i|\Phi\rangle}\leq0\ ,
\label{eq:genmulti}
\ee
\noindent with
${\cal P}_i=\Pi_{A_i}\otimes\id_{B_i}$, and where the sum runs over all inequivalent bipartitions.\\
To convince ourselves that ineq.~\eqref{eq:genmulti} is indeed satisfied by all at least partially separable states $\rho$,
let us first verify that this holds for any {\em pure} state $\rho_\Psi=\ket{\Psi_{bs}}\bra{\Psi_{bs}}$
that is biseparable with respect some partition labeled $i_0$.
Just like any duplicated state $\ket{\Psi}^{\otimes 2}$ is invariant under the global permutation ${\bf \Pi}$,
the duplicated state $\ket{\Psi_{bs}}^{\otimes 2}$ is invariant under $\Pi_{A_{i_0}}\otimes\id_{B_{i_0}}$.
Therefore, the first term in ineq.~\eqref{eq:genmulti} cancels with the $i=i_0$ term in the summation.
All remaining terms are expectation values of positive operators, and given the negative sign in front of the sum,
the left-hand-side is indeed nonpositive. 
Hence, ineq.~\eqref{eq:genmulti} is satisfied for any pure state that is not genuinely multipartite entangled.

The generalization of ineq.~\eqref{eq:genmulti} to mixed states is a direct consequence of its convexity which we can see in the following,
where we will use that the state $\ket{\Phi}$ is completely separable.
That is, independently of which decomposition of the Hilbert space into two subspaces we take,
we can always write it as a direct product of two states $\ket{\Phi_1}$ and $\ket{\Phi_2}$ of the respective subspaces. 
The first term in ineq.~\eqref{eq:genmulti} is the absolute value of
the matrix element $\bra{\Phi_1}\rho\ket{\Phi_2}$:
\beq
\sqrt{\langle\Phi|\rho^{\otimes 2}{\bf \Pi}|\Phi\rangle}
=|\bra{\Phi_1}\rho\ket{\Phi_2}|\ ,
\label{eq:absdiag}
\eeq
since ${\bf \Pi}$ simply permutes $\ket{\Phi_1}$ and $\ket{\Phi_2}$,
{\it i.e.} ${\bf \Pi}\ket{\Phi_1}\otimes\ket{\Phi_2}=\ket{\Phi_2}\otimes\ket{\Phi_1}$.
And the absolute value is convex, {\it i.e.} $|a+b|\le|a|+|b|$ for arbitrary complex numbers $a$ and $b$.
Each summand
${\cal K}_i=\sqrt{\langle\Phi|{\cal P}_i^\dagger\rho^{\otimes 2}{\cal P}_i|\Phi\rangle}$
in the second term of ineq.~\eqref{eq:genmulti} is the square root of a product of two diagonal density matrix elements,
{\it i.e.} non-negative numbers
\beq
{\cal K}_i=\sqrt{\bra{\widetilde{\Phi}_1}\rho\ket{\widetilde{\Phi}_1}\bra{\widetilde{\Phi}_2}\rho\ket{\widetilde{\Phi}_2}}\ ,
\label{eq:second_term}
\eeq
with $\ket{\widetilde{\Phi}_1}\otimes\ket{\widetilde{\Phi}_2}=\Pi_{A_i}\otimes\id^{B_i}\ket{\Phi}$.
Now, Cauchy-Schwarz's inequality $\sum_jp_jq_j\ge\sqrt{\sum_jp_j^2}\sqrt{\sum_jq_j^2}$ with
$p_j=\sqrt{\bra{\widetilde{\Phi}_1}\rho_j\ket{\widetilde{\Phi}_1}}$ and $q_j=\sqrt{\bra{\widetilde{\Phi}_2}\rho_j\ket{\widetilde{\Phi}_2}}$
yields
\be
{\cal K}_i\le\sum_j\sqrt{\bra{\widetilde{\Phi}_1}\rho_j\ket{\widetilde{\Phi}_1}\bra{\widetilde{\Phi}_2}\rho_j\ket{\widetilde{\Phi}_2}}\ ,
\ee
for any set of positive operators $\rho_j$ satisfying $\rho=\sum_j\rho_j$.
Therefore, Eq.~\eqref{eq:second_term} is a concave quantity, so that ineq.~\eqref{eq:genmulti} is indeed convex.
Since, as shown above, it is satisfied for all biseparable {\em pure} states, this implies the same also for {\em mixed} states.

Ineqs.  \eqref{eq:bilin} and \eqref{eq:genmulti} are valid for any choice of a completely separable pure state-vector $\ket{\Phi}$,
but the potential to detect the genuine multipartite character of a given entangled state will depend on a suitable choice of $\ket{\Phi}$.
For a state with rather weak genuine multipartite entanglement an optimization, {\it i.e.}
a search for the product-vector $\ket{\Phi}$ that maximizes the violation of the respective inequality might be necessary.
Optimizations that are quite common in the theory of entangled states, and generally pose a difficult problem.
But, finding an optimal state vector is significantly easier than
{\it e.g.} convex roof construction \cite{PhysRevA.62.052310}, or optimization of entanglement witnesses \cite{PhysRevA.62.032307}
since there are no constraints to be satisfied
and efficient algorithms are available \cite{shh}.
However, what is even more important, the optimization space grows only linearly with the number of subsystems as opposed to the typically exponential scaling of such problems.
Besides such an optimization, one can combine different choices of states $\ket{\Phi}$
to tailor criteria that are suited particularly well for a specific class of states, as we demonstrate here with the exemplary
choice of $|\Phi_{ij}\rangle=|s_i\rangle\otimes|s_j\rangle$ with
$\ket{s_i}=|x\dots xyx\dots x\rangle$ in terms of two single-particle states $\ket{x}$ and $\ket{y}$,
and $\ket{y}$ is chosen exactly for the $i$-th entry of $\ket{s_i}$. 
Taking linear combinations of ineq.~\eqref{eq:genmulti} for these choices we arrive at
\begin{align}\tag{III}
\sum_{i\neq j} \sqrt{\langle \Phi_{ij}|\rho^{\otimes 2}{\bf \Pi}|\Phi_{ij}\rangle}-
(n-2)\sum_{ij} \sqrt{\langle  \Phi_{ij}|{\cal P}_i^\dagger\rho^{\otimes 2}{\cal P}_i| \Phi_{ij}\rangle} \leq 0 
\label{eq:genmulti2}
\end{align}
where ${\cal P}_i=\Pi_{A_i}\otimes\id_{B_i}$ is defined analogously to the above.
However, in contrast to the above, not all bipartitions are taken into account, but $A_i$ is the duplicated Hilbert space of the $i$-th subsystem
and $B_i$ the rest.
Exactly as in ineq.~\eqref{eq:genmulti} also the left-hand-side in ineq.~\eqref{eq:genmulti2} is convex,
so that the inequality is proven for biseparable {\em mixed} states,
since it is proven for biseparable {\em pure} states in the end of the paper.
\begin{figure}[t]
\begin{center}
\small{(a)}
\includegraphics[width=3.7cm, keepaspectratio=true]{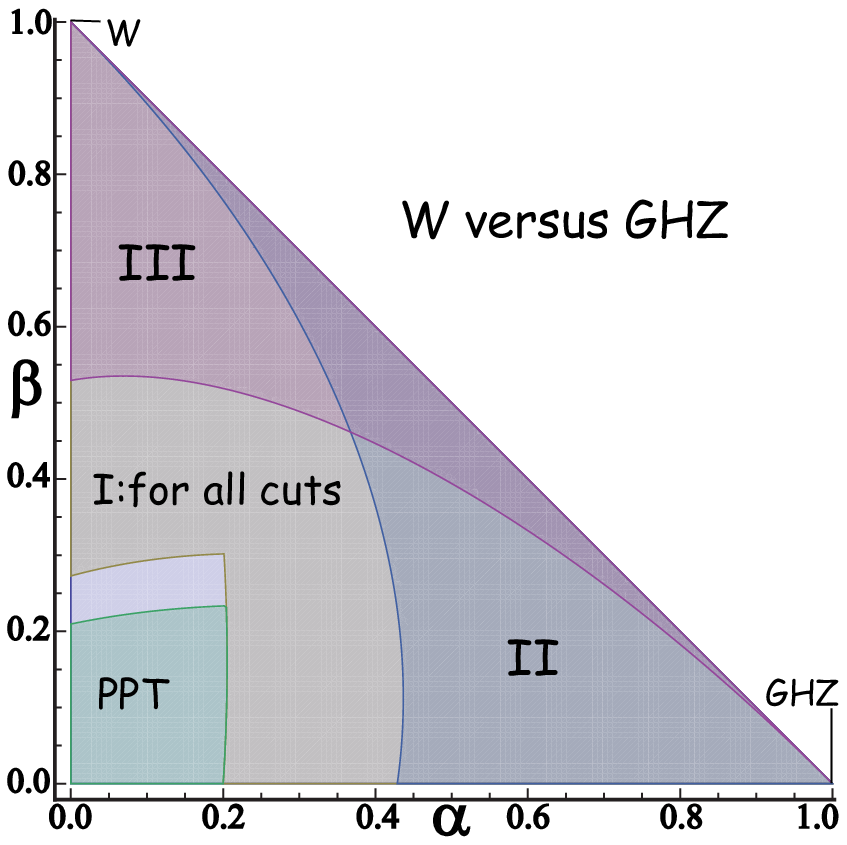}
\small{(b)}
\includegraphics[width=3.7cm, keepaspectratio=true]{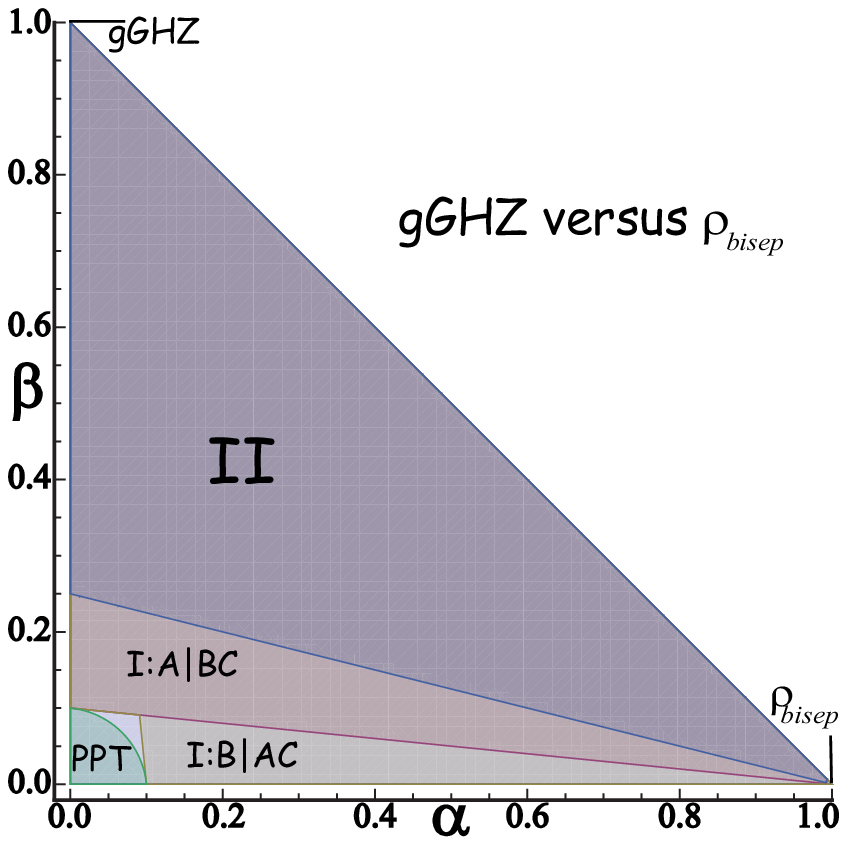}
\caption{Here the detection quality of the bilinear inequalities
(\ref{eq:bilin}), (\ref{eq:genmulti}) and (\ref{eq:genmulti2}) is
shown for the state
$\rho=\frac{1-\alpha-\beta}{8}\id+\alpha\rho_{GHZ}+\beta \rho_W$ (a) and the tripartite qutrit state (with subsystems labeled $ABC$)
$\rho=\frac{1-\alpha-\beta}{27}\id+\alpha\rho_{bisep}+\beta\rho_{gGHZ}$ (b).
Area \ref{eq:genmulti} contains genuine multipartite entanglement
detected by (\ref{eq:genmulti}). Area \ref{eq:genmulti2} contains genuine multipartite
entanglement detected by (\ref{eq:genmulti2}). Area \ref{eq:bilin} (a) is not biseparable w.r.t. any
bipartition, since it violates inequality (\ref{eq:bilin}) for all
partitions, area \ref{eq:bilin}:$B|AC$ is not biseparable w.r.t. $B|AC$, since it violates
inequality (\ref{eq:bilin}) for this partition (the result for
$AB|C$ is equivalent). Area \ref{eq:bilin}:$A|BC$ contains states that
violate inequality (\ref{eq:bilin}) for bipartition $A|BC$. The area labeled PPT constitutes all states not detected by the Peres-Horodecki criterion \cite{horodeckiqe}.}\label{qubit}
\end{center}
\end{figure}

In particular with growing system size the ability to assess separability criteria efficiently is getting more and more important
since quantum state tomography (QST) scales so unfavorably with the number of subsystems.
Being bilinear expectation values, the present criteria can very efficiently be measured experimentally on identically prepared quantum states as it has been done in \cite{{bilin}}.
But also with measurements performed on individually prepared quantum states the present criteria can be experimentally assessed with significantly fewer observables than required for QST:
Eq.~\eqref{eq:genmulti} is given in terms the square root of the number of observables needed for QST and Eq.~\eqref{eq:genmulti2} scales as $2n(n-1)$,
{\it i.e.} polynomially with the number $n$ of subsystems.
 
An indispensable prerequisite for any practical criterion is its robustness against experimental imperfections.
Therefore, let us discuss the capacity of our present criteria on a few exemplary quantum states,
where we test this robustness.\\
\emph{Example 1.} First consider the three qubit state
$\rho=\frac{1-\alpha-\beta}{8}\id+\alpha\,\rho_{GHZ}+\beta\,\rho_{W}$
where
$\rho_{GHZ}=\frac{1}{2}(|000\rangle+|111\rangle)(\langle000|+\langle111|)$
and
$\rho_{W}=\frac{1}{3}(|001\rangle+|010\rangle+|100\rangle)(\langle001|+\langle010|+\langle100|)$.
It is a mixture of the $GHZ$-state and the $W$-state dampened
by isotropic noise (see Ref.\cite{acin} for further details). In Fig.~\ref{qubit}a the detection parameter spaces of the inequalities (\ref{eq:bilin}), (\ref{eq:genmulti}) and (\ref{eq:genmulti2}) are illustrated.
In the case of genuine multipartite entanglement detection for
qubits, these criteria work as well as the best known method so far.
For example in Ref.~\cite{guehnewit} the above state for
($\alpha=0$ and $\beta=1-p$) was found to be genuinely multipartite
entangled by means of entanglement witnesses up to a threshold of $p
< 8/19$. This bound was then improved to $p < 8/17$ \cite{guehnecrit},
which is also our result. In fact for this special case our criteria coincide. For qudits, our criteria are the first detection criteria known so far.\\
\emph{Example 2.} Consider the three qutrit state
$\rho=\frac{1-\alpha-\beta}{27}\id+\alpha\rho_{bisep}+\beta\rho_{gGHZ}$ where
$\rho_{gGHZ}=\frac{1}{3}(|000\rangle+|111\rangle+|222\rangle)(\langle000|+\langle111|+\langle222|)$
and
$\rho_{bisep}=\frac{1}{2}(|0\rangle\langle0|\otimes(|00\rangle+|11\rangle+|22\rangle)(\langle00|+\langle11|+\langle22|)$.
It is a mixture between a generalized $GHZ$-state for qutrits and a
biseparable qutrit state dampened by isotropic noise. In
Fig.~\ref{qubit}b the detection parameter spaces of the violation of
the inequalities (\ref{eq:bilin}) and (\ref{eq:genmulti}) are
illustrated.\\
\emph{Example 3.} Now consider the following four qudit state:
\be
\rho_S=\frac{1-\alpha-\beta}{d^4}\id+\frac{\alpha}{d}\sum_i\rho^i_{gGHZ1}+\frac{\beta}{d}\sum_i\rho^i_{gGHZ2}
\ee
where $\rho^i_{gGHZx}:=|gGHZx(i)\ra\la gGHZx(i)|$ with $|gGHZx(i)\ra:=\sum_k\frac{1}{\sqrt{d}}|k\ra|k\oplus x\ra|k\oplus i\ra|k\oplus i\oplus x\ra$. Where $\oplus$ is the addition modulo $d$. For $d=2$ and $\alpha=\beta$ this is the bound entangled Smolin state (see Ref.~\cite{smolin}) dampened by isotropic noise.
Also in this case our criteria work well. Ineq. (\ref{eq:bipartite}) shows that all states in the region $1-(d^2 +1)\alpha-\beta < 0$ and $1-\alpha-(d^2 +1)\beta < 0$ are not separable with respect to any bipartition.
Moreover ineq. (\ref{eq:genmulti}) shows that the state only becomes biseparable outside the region not detected by ineq. (\ref{eq:bipartite}) for $\alpha = \beta > 1/(d^2 + 2)$, i.e. all entangled states in this region except for the line $\alpha = \beta$ are definitely multipartite entangled.
That is, our criteria detect all states that are detected by  the Peres-Horodecki criterion ~\cite{horodeckiqe},
but, -- exceeding the scope of the latter --  characterize states to be genuinely multipartite entangled.\\
\emph{Example 4}
Consider the generalized $GHZ$ state
$|\psi_{dn}\rangle=\frac{1}{\sqrt{d}}\sum_{i=0}^{d-1}|i\rangle^{\otimes n}$
with additional
isotropic (white) noise:
\be
\rho=p|\psi_{dn}\rangle\langle\psi_{dn}|+(1-p)\frac{1}{d^n}\id
\ee
With inequality (\ref{eq:genmulti}) we can show analytically that these states are genuinely multipartite entangled for $p>\frac{3}{d^{n-1}+3}$, which shows that even in high dimensional systems with many constituents these criteria work very well.

In conclusion ineq.~\eqref{eq:genmulti2} is only one specific of many possible criteria derived from ineq.~\eqref{eq:genmulti} and the versatility of our approach allows to tailor many criteria
suited for specific classes of entangled states.
Given the efficient decomposition in physical observables, our criteria enable verification of mixed state entanglement
with tools \cite{PhysRevA.74.022314,arXiv:1003.4862} originally only applicable to pure states.

\noindent\emph{Appendix}: Finally, let us prove ineqs.~\eqref{eq:bipartite} and \eqref{eq:genmulti2}.
For the former, we have to show
\beq
\bra{\Phi}\varrho_s^{\otimes m}\ket{\Phi}\ge
\frac{1}{2}(\bra{\Phi}{\cal P}_A^\dagger\varrho_s^{\otimes m}{\cal P}_B\ket{\Phi}+\bra{\Phi}{\cal P}_B^\dagger\varrho_s^{\otimes m}{\cal P}_A\ket{\Phi})\nonumber
\label{eq:toshow}
\eeq
for any separable mixed state
$\varrho_s=\sum_i \ket{\varphi_i}\bra{\varphi_i}\otimes\ket{\chi_i}\bra{\chi_i}$
and any completely separable state-vector
$\ket{\Phi}=\bigotimes_{i=1}^m\ket{\alpha_i}\otimes\bigotimes_{i=1}^m\ket{\beta_i}$.
This amounts to showing
\be
\vec X^\ast\vec X-\frac{1}{2}\left(\Pi_A\vec X\right)^\ast \left(\Pi_B\vec X\right)-\frac{1}{2}\left(\Pi_B\vec X\right)^\ast \left(\Pi_A\vec X\right)\ge 0\ ,
\label{eq:PiX}
\ee
with
$[\vec X]_{p_1\hdots p_mq_1\hdots q_n}=\prod_{i=1}^m\langle\alpha_i|\varphi_{p_i}\rangle\prod_{i=1}^m\langle\beta_i|\chi_{q_i}\rangle$, $[\Pi_A\vec X]_{p_1\hdots p_mq_1\hdots q_n}=\prod_{i=1}^m\langle\alpha_i|\varphi_{p_{i+1\mbox{mod}m}}\rangle\prod_{i=1}^m\langle\beta_i|\chi_{q_i}\rangle$,
$[\Pi_B\vec X]_{p_1\hdots p_mq_1\hdots q_n}=\prod_{i=1}^m\langle\alpha_i|\varphi_{p_i}\rangle\prod_{i=1}^m\langle\beta_i|\chi_{q_{i+1\mbox{mod}m}}\rangle$.
Since $\left(\Pi_{A/B}\vec X\right)^\ast\left(\Pi_{A/B}\vec X\right)=\vec X^\ast\vec X$, ineq.~\eqref{eq:PiX} simplifies to
$\frac{1}{2}\left|\Pi_A\vec X-\Pi_B\vec X\right|^2\ge 0$,
which proves ineq.\eqref{eq:bipartite}.
Similar to the proof presented in Ref. ~\cite{guehnecrit} we only have to verify that ineq.~\eqref{eq:genmulti2} is satisfied for pure biseparable states $\ket{\Psi}$ due to its convexity (as shown for ineq.~\eqref{eq:genmulti}). With the short hand notation
$x_{ij}=\sqrt{\langle \Phi_{ij}|\rho_\Psi^{\otimes 2}{\bf \Pi}|\Phi_{ij}\rangle}$ and
$y_{ij}=\sqrt{\langle  \Phi_{ij}|{\cal P}_i^\dagger\rho_\Psi^{\otimes 2}{\cal P}_i| \Phi_{ij}\rangle}$,
ineq.~\eqref{eq:genmulti2} reads
$\sum_{i\neq j}x_{ij}-(n-2)\sum_{ij}y_{ij}$.
We will have to distinguish between the cases in which both indices $i$ and $j$ correspond to different, or the same parts $A$ and $B$ in the bipartition with respect to which $\ket{\Psi}$
(without loss of generality we assume $i$ to correspond to $A$).
The former contributions to ineq.~\eqref{eq:genmulti2} we denote as
$B_d=\sum_{i,j\in B} (x_{ij} -(n-2)y_{ij})$, the latter as
$B_s=\sum_{i\neq j\in A} (x_{ij} -(n-2)y_{ij})-(n-2) \sum_i y_{ii}$, so that ineq.~\eqref{eq:genmulti2}  reads $B_s+B_d\le 0$.
$B_d$ is non-positive since $x_{ij}\le y_{ij}$ as shown for ineq.~\eqref{eq:genmulti}.
Since the $y_{ij}$ are non-negative,
we obtain
$B_s\le\sum_{i\neq j\in s}x_{ij}-(n-2)\sum_i y_{ii}=\sum_{i\neq j\in s}(x_{ij}-z_iy_{ii})\le\sum_{i\neq j\in s}(x_{ij}-y_{ii})$
with $z_i=(n-2)/(n_i-1)$, where $n_i$ is the number of subsystems in $A$,
where $z_i\ge 1$ since $A$ can comprise at maximum $n-1$ subsystems.
Now, we can symmetrize the last term in $\sum_{i\neq j\in s}(x_{ij}-y_{ii})$, {\it i.e.} rewrite it as
$\sum_{i\neq j\in s}(x_{ij}-1/2(y_{ii}+y_{jj}))$.
Since $y_{ii}=\bra{s_i}\varrho_\Psi\ket{s_i}$ (due to the relation ${\cal P}_i\ket{\Phi_{ii}}=\ket{\Phi_{ii}}$),
we can conclude $x_{ij}=|\bra{s_i}\varrho_\Psi\ket{s_i}|\le1/2(y_{ii}+y_{jj}))$,
such that $B_s$ is non-negative, what finishes the proof of ineq.~\eqref{eq:genmulti2}.\\
\noindent\emph{Acknowledgements.}
We would like to thank T. Adaktylos, H. Schimpf and C.
Spengler for productive discussions.
M. Huber and A. Gabriel gratefully acknowledge the Austrian Fund project FWF-P21947N16 and
F. Mintert funding within the DFG project MI1345/2-1.

\end{document}